\documentclass[12pt]{article}

\title{Classical and quantum spaces as rough images of the fundamental
prespace}
\author{Andrei Khrennikov\footnote{International Center for Mathematical Modeling 
in Physics and Cognitive Sciences, Andrei.Khrennikov@msi.vxu.se; supported by EU-Network
 ``QP and Applications.''} \\
MSI, University of V\"axj\"o, S-35195, Sweden}
\begin{document}
\maketitle

\begin{abstract} In spite of all {\bf no-go} theorems (e.g., von Neumann, Kochen and Specker,...,
Bell,...) we constructed a realist basis of quantum mechanics. In our model both
classical and quantum spaces b are rough images of the fundamental
{\bf prespace.} Quantum mechanics cannot be reduced to classical one. Both classical and quantum
representations induce reductions of prespace information.
\end{abstract}

{\bf 1. Introduction.} In preprints [1] \footnote{ Plenary talk and a topic of the round table 
at the International Conference
``Quantum Theory: Reconsideration of Foundations-2'', June-2003, V\"axj\"o, Sweden.}
there was constructed a contextual quantum representation of the Kolmogorovian model.
That mathematical construction can be used as a {\bf realist basis of quantum 
mechanics} (QM). Existence of such a realist ``underground" of QM was the question 
of the great debate since first days of QM, see, e.g., [2], [3] for detail. It should
be reminded that A. Einstein strongly supported the idea that such a realist underground
of QM could finally be found and W. Heisenberg and N. Bohr claimed that it would be
impossible. In this note we present main ideas of [1] without to go in rather technical mathematical
details of contextual representation of the Kolmogorovian model in a Hilbert space.

It should be underlined from the very beginning that we do not discuss a reduction of quantum physics
to classical one.

{\bf 2. Prespace, classical space and quantum space.} In my model both quantum and
classical states are rough images of contexts --
complexes of physical conditions. In the mathematical model  [1], cf. 
[4], contexts are described as sets of fundamental parameters. We call the space of fundamental
parameters {\bf prespace} and denote it $\Omega.$ Contexts  are 
representated by a family of subsets of $\Omega.$
The prespace  $\Omega$ is underground of the
classical space  $X_{\rm{cl}}={\bf R}^3$ as well as the quantum (Hilbert) space 
$X_{\rm{q}}=H.$  QM gives essentially
richer picture of the prespace $\Omega:$ the QM-representation of $\Omega$-contexts generates essentially
larger class of images than the classical representation. In particular, it is impossible to reduce
quantum picture of the prespace $\Omega$ to the classical one.

This is a very delicate point of considerations. Dynamics in the prespace $\Omega$ is a deterministic
dynamics. But it is not ``classical dynamics'' since the latter takes place not in the prespace
$\Omega$ but in the classical space $X_{\rm{cl}}.$ This is not a question of the mathematical
realization of the prespace $\Omega$ and the classical space $X_{\rm{cl}}.$ It may be that the prespace
$\Omega$ can also be described as $\Omega= {\bf R}^m$ (or even as $\Omega= {\bf R}^3$ -- so in the same way
as $X_{\rm{cl}}={\bf R}^3,$ cf. section 8). The crucial point is that $X_{\rm{cl}}$ is created via the huge
{\bf reduction of information} in the process of transition from the prespace contexts to points of 
$X_{\rm{cl}}.$ Each classical point $x \in X_{\rm{cl}}$ is the image of a domain $B_x$ of 
the prespace $\Omega$, see [1], and this domain can contain huge (may be even infinite) number 
of prepoints $\omega =\omega^x.$
By our model there exist mappings:
$$
\Omega \to X_{\rm{cl}}, \; \; \; \Omega \to X_{\rm{q}}, \; \; \;\mbox{and}\; \; \;
 X_{\rm{q}}\to X_{\rm{cl}}
$$
Thus we also obtain the map:
$$
 \Omega \to X_{\rm{q}}\to X_{\rm{cl}}
$$
But there is no pathway:
$$
\Omega \to X_{\rm{cl}}\to X_{\rm{q}}
$$
We underline (see further considerations) that the correspondence principle is based on the
map $\Omega \to X_{\rm{q}}$ and not at all on the map $X_{\rm{q}}\to X_{\rm{cl}}.$

{\bf 3. Fundamental incompatible preobservables.} Contextual probabilistic representation [1] of 
$\Omega$-contexts in the quantum space $X_{\rm{q}}=H$  is based on a fixed pair of incompatible 
preobservables  ({\bf reference observables}):
\begin{equation}
\label{RO}
b, a: \Omega \to {\bf R}
\end{equation}
In our model [1] preobservables are functions $d: \Omega \to {\bf R}.$ Denote the set of all
preobservables ${\cal O} (\Omega).$ We interpret observables $d \in  {\cal O} (\Omega)$
as {\bf realist observables:} by fixing a prepoint $\omega \in \Omega$ we are able to fix the value
$d= d(\omega).$
We are not able to measure an arbitrary $d \in  {\cal O} (\Omega).$
But reference preobservables (\ref{RO}) and functions of those preobservables,
$f(b), f(a)$ can be measured [1].
We denote the space of quantum observables by the symbol
${\cal O}(H).$ In mathematical models ${\cal O}(\Omega)$ and ${\cal O}(H)$ are represented by spaces
of (Kolmogorovian) random variables and self-adjoint operators, respectively.

{\bf 4. Position-momentum picture of prespace.} In principle a contextual probabilistic representation of the prespace $\Omega$
in the quantum space $X_{\rm{q}}=H$ can be based on any pair of incompatible preobservables. However, it seems
that we (human beings) can use only the special pair of reference preobservables:
$$
(q,p)= (\rm{position, momentum}).
$$
Thus modern quantum as well as classical physics are the {\it position-momentum pictures} of the 
prespace. All  classical and quantum  observables are functions of position and momentum
observables. In quantum case we use functions $\hat{d}= u(\hat{q}, \hat{p})$ of 
operators  of the position $\hat{q}$ and the momentum $\hat{p}.$

By choosing another pair of reference preobservables we obtain another quantum picture
of the prespace $\Omega.$ However, it seems that at the present time we are not able to measure
preobservables $d \in  {\cal O} (\Omega)$ distinct from functions $f(q)$ and $g(p).$

{\bf 5. Nonequivalence of quantum pictures of the prespace.} It should be underlined that 
quantum pictures of the prespace $\Omega$ based on two different pairs of incompatible observables
$(b,a)$ and $(v,w)$ are in general nonequivalent. Of course, the same mathematical formalism --
the Hilbert space formalism -- can be used for any quantum picture of the $\Omega.$\footnote{But
Hilbert spaces $H^{b/a}$ and $H^{v/w}$ corresponding to pictures based on 
$(b,a)$ and $(v,w)$  can be different.} But we should pay attention to physical structures of representations.
So we should not forget about the $(q,p)$-origin of  QM (as a physical theory and not as only a mathematical
formalism), see also [5], [6].

{\bf 6. No ``no-go?''} Existence of a realist underground model of QM looks very surprising in the 
view of various no-go theorems, e.g., von Neumann [7], Kochen-Specker [8], Bell [9],...
But all those no-go theorems suffered of the absence of physical justification for the list of assumptions
on the correspondence between a realist prequantum model and QM. 
J. Bell performed the brilliant analysis of assumptions on 
the ``real-quantum" correspondence
which were assumed (very often indirectly) in previous no-go theorems, see [9]. We should agree with Bell
that von Neumann, Kochen and Specker and many others wanted too much for 
the ``real-quantum" correspondence. Thus despite all pre-Bellian no-go theorems J. Bell was sure that
it is possible to construct a realist basis of QM. However, J. Bell also wanted too much for
the ``real-quantum" correspondence, see,e.g., [10], [6] for analysis of Bell's assumptions. 
As a consequence, he came to the conclusion that every realist prequantum model should be
{\bf nonlocal.}

{\bf 7. Correspondence between preobservables and quantum observables.} Correspondence-maps
$$
W: {\cal O}(\Omega)  \to {\cal O}(H)
$$
between realist preobservables and quantum observables which were considered by von Neumann,
Kochen and Specker, ..., Bell,... were too straightforward. Neither von Neumann and Kochen-Specker
nor Bell had physical arguments to present a list of ``natural features'' of such a correspondence 
$W.$ I neither have physical arguments. But I have strong probabilistic arguments. There
exists a unique quantum representation of a Kolmogorovian model and this representation automatically
induces a map $W$ which have very special properties [1]. Neither von Neumann and Kochen-Specker
nor Bell maps have such properties. 

In our realist model the map $W$ is  defined only on a proper domain $D_W$ in 
${\cal O}(\Omega),$ namely
$$
D_W={\cal O} (b,a)=\{d(\omega) = f(b(\omega)) + g(a(\omega))\},
$$
where $(b,a)$ is the pair of reference preobservables determining the quantum picture of the $\Omega.$
And in general the map $W$ does not preserve conditional probability distributions. Here we consider
two conditionings: contextual conditioning in the $\Omega$ and quantum state coditioning 
in the $H.$ 

But (!) conditional averages are preserved by the map $W:$
$$
E(d/C)= (\hat{d}\phi_C, \phi_C),
$$
where $\hat{d}= W(d), d\in D_W, $ and $\phi_C\in H$ is the image of a prespace context $C.$

It is very important that quantum Hamiltonians belong to the $W$-image of the set
of  preobservables. The operator
$$
\hat{{\cal H}}= \frac{\hat{a}^2}{2} + V(\hat{b})
$$
is the image of the  energy preobservable 
$$
{\cal H}(\omega)= \frac{a(\omega)^2}{2} + V(b(\omega)).
$$
But as we have already underlined the $W$ does not preserve probability distributions.
So $\hat{{\cal H}}$ and ${\cal H}(\omega)$ have different probability distributions.
But they have the same average. 

{\bf 8. Quantization and the correspondence principle.} As we noticed, it was proved [1]
that quantum Hamiltonian $\hat{{\cal H}}$ has the same contextual average as the
prespace Hamiltonian ${\cal H}:$ 
\begin{equation}
\label{CPR}
E({\cal H}/C)= (\hat{{\cal H}}\phi_C, \phi_C).
\end{equation}
We can speculate that this coincidence of averages is the real basis of quantization
rules. By (\ref{CPR}) to obtain the correct average of the energy preobservable
${\cal H}(\omega)$ we should put quantum images of the reference preobservables into 
the prespace Hamiltonian:
\begin{equation}
\label{CPR1}
b\to \hat{b}=W(b), \; a\to \hat{a}=W(a)
\end{equation}
and, in particular, for the (position, momentum) quantization
\begin{equation}
\label{CPR2}
q_\Omega\to \hat{q}=W(q_\Omega), \; p_\Omega\to \hat{p}=W(q_\Omega)
\end{equation}

We should sharply distinguish the prespace position and momentum, 
$q_\Omega, p_\Omega,$ and classical space position and momentum 
$q_{X_{\rm{cl}}}, p_{X_{\rm{cl}}}.$ In our model (in the opposite 
to the very common opinion) $\hat{q}\not=W(q_{X_{\rm{cl}}})$ 
and $\hat{p}\not=W(p_{X_{\rm{cl}}}).$ 

By our model the root of the quantization rule is the equality 
(\ref{CPR}) and Hamiltonian dynamics in the prespace $\Omega$
and not Hamiltonian dynamics in the classical space $X_{\rm{cl}}.$
\footnote{It may be that dynamics in $\Omega$ and $X_{\rm{cl}}$ 
are mathematically described in the same way. But we should distinguish
Hamiltonian prespace dynamics and classical Hamiltonian dynamics.}

{\bf Conclusion.} {\it In spite of all {\bf no-go} theorems the realist
model of QM exists.}

I would like to thank A. Plotnitsky for numerous discussions on 
Heisenberg-Bohr interpretation of QM and L. Accardi, L. Ballentine, 
S. Gudder for discussions on the role of conditional probabilities
in QM which were extremely important for the creation of a contextual
representation of a Kolmogorovian model in a Hilbert space [1], cf. [4].

{\bf References}

1. A. Khrennikov, {\it Contextual approach to quantum mechanics and the theory
of the fundamental prespace.} quant-ph/0306003.

2. A. Plotnitsky,  Quantum atomicity and quantum information: Bohr, Heisenberg,
and quantum mechanics as an information theory.
Proc. Int. Conf. {\it Quantum Theory: Reconsideration
of Foundations.} Ed. A. Khrennikov, Ser. Math. Modelling, {\bf 2},
309-342, V\"axj\"o Univ. Press, 2002\\
(http://www.msi.vxu.se/forskn/quantum.pdf).

3. A. Khrennikov, {V\"axj\"o interpretation of quantum mechanics.}
Proc. Int. Conf. {\it Quantum Theory: Reconsideration
of Foundations.} Ed. A. Khrennikov, Ser. Math. Modelling, {\bf 2}, 164-169,
V\"axj\"o Univ. Press, 2002 (http://www.msi.vxu.se/forskn/quantum.pdf).

4. A. Yu. Khrennikov, Ensemble fluctuations and the origin of quantum probabilistic rule.
{\it J. Math. Phys.}, {\bf 43}, N. 2, 789-802 (2002).

5. L. De Broglie, {\it The current interpretation of quantum mechanics. A critical
study.} Elsevier Publ., Amsterdam (1964).

6. A. Yu. Khrennikov, I. V. Volovich, Local Realism, Contextualism  and
Loopholes in Bell`s Experiments.  Proc. Int. Conf. {\it Foundations of Probability and Physics-2.}
Ed. A. Khrennikov, Ser. Math. Modelling, {\bf 5}, 325-344, V\"axj\"o Univ. Press, 2003. 
 (quant-ph/0212127).

7. J. von Neumann, {\it Mathematical foundations
of quantum mechanics.} Princeton Univ. Press, Princeton, N.J. (1955).

8.  S. Kochen and E. Specker, {\it J. Math. Mech.}, {\bf 17}, 59-87 (1967).

9. J. S. Bell, {\it Speakable and unspeakable in quantum mechanics.} Cambridge Univ. Press (1987).

10.  Khrennikov A.Yu., {\it Interpretations of Probability.}
VSP Int. Sc. Publishers, Utrecht/Tokyo (1999).

\end{document}